\documentclass[aps,prl,twocolumn,showpacs,groupedaddress]{revtex4}  % for review and submission 
\usepackage{graphicx}  % needed for figures
\usepackage{dcolumn}   % needed for some tables
\usepackage{bm}        % for math
\usepackage{amssymb}   % for math

\usepackage{epsfig}
\usepackage{cancel}

\newcommand{\ppbar}{\ensuremath{p\bar{p}}}
\newcommand{\tanb}{\ensuremath{\tan\beta}}
\newcommand{\cha}{\ensuremath{\tilde{\chi}^{\pm}}}
\newcommand{\chaone}{\ensuremath{\tilde{\chi}^{\pm}_1}}
\newcommand{\chiz}{\ensuremath{\tilde{\chi}^0}}
\newcommand{\chizone}{\ensuremath{\tilde{\chi}^0_1}}
\newcommand{\chiztwo}{\ensuremath{\tilde{\chi}^0_2}}

\newcommand{\mcha}{\ensuremath{m_{\tilde{\chi}^{\pm}_1}}}
\newcommand{\mchione}{\ensuremath{m_{\tilde{\chi}^0_1}}}
\newcommand{\mchitwo}{\ensuremath{m_{\tilde{\chi}^0_2}}}

\newcommand{\mzero}{\ensuremath{m_0}}
\newcommand{\mhalf}{\ensuremath{m_{1/2}}}
\newcommand{\mmplane}{\mzero--\mhalf\ plane}

\newcommand{\met}{\mbox{\ensuremath{\not \hspace{-1.0mm} E_T}}}

\newcommand{\dphi}{\ensuremath{\Delta\phi_{\ell_{1}\ell_{2}}}}

\newcommand{\fbinv}{\ensuremath{\mathrm{fb^{-1}}}}
\newcommand{\ttype}{\ensuremath{\tau}-type}

\newcommand{\ptone}{\ensuremath{p_T^{\ell1}}}
\newcommand{\pttwo}{\ensuremath{p_T^{\ell2}}}

\newcommand{\pttr}{\ensuremath{p_T^{tr}}}
\newcommand{\mll}{\ensuremath{m_{\ell_{1\!,2},tr}}}

\newcommand{\sigbr}{\ensuremath{\sigma\times\mathrm{BR}(3\ell)}}

\newcommand{\eel}{\ensuremath{ee\ell}}
\newcommand{\mumul}{\ensuremath{\mu\mu \ell}}
\newcommand{\mutaul}{\ensuremath{\mu\tau \ell}}
\newcommand{\mutau}{\ensuremath{\mu\tau}}
\newcommand{\emul}{\ensuremath{e\mu \ell}}

\begin{document}

% the following line is for submission, including submission to the arXiv!!
\hspace{5.2in} \mbox{Fermilab-Pub-09/003-E}

\title{Search for associated production of charginos and neutralinos in
  the trilepton final state using 2.3~fb{\boldmath $^{-1}$} of data}
% LIST_OF_AUTHORS_R2.TEX                 11/25/08           
%
\author{V.M.~Abazov$^{36}$}
\author{B.~Abbott$^{75}$}
\author{M.~Abolins$^{65}$}
\author{B.S.~Acharya$^{29}$}
\author{M.~Adams$^{51}$}
\author{T.~Adams$^{49}$}
\author{E.~Aguilo$^{6}$}
\author{M.~Ahsan$^{59}$}
\author{G.D.~Alexeev$^{36}$}
\author{G.~Alkhazov$^{40}$}
% ...

\affiliation{\vspace{0.1 in}(The D\O\ Collaboration)\vspace{0.1 in}}
% ...
\affiliation{$^{6}$University of Alberta, Edmonton, Alberta, Canada,
                Simon Fraser University, Burnaby, British Columbia, Canada,
                York University, Toronto, Ontario, Canada, and
                McGill University, Montreal, Quebec, Canada}
% ...
\affiliation{$^{29}$Tata Institute of Fundamental Research, Mumbai, India}
% ...
\affiliation{$^{36}$Joint Institute for Nuclear Research, Dubna, Russia}
% ...
\affiliation{$^{40}$Petersburg Nuclear Physics Institute,
                St. Petersburg, Russia}
% ...
\affiliation{$^{49}$Florida State University, Tallahassee, Florida 32306, USA}
% ...
\affiliation{$^{51}$University of Illinois at Chicago,
                Chicago, Illinois 60607, USA}
% ...
\affiliation{$^{59}$Kansas State University, Manhattan, Kansas 66506, USA}
% ...
\affiliation{$^{65}$Michigan State University,
                East Lansing, Michigan 48824, USA}
% ...
\affiliation{$^{75}$University of Oklahoma, Norman, Oklahoma 73019, USA}
% ...
  % input Dzero author list
\date{January 6, 2009}

\begin{abstract}
We report the results of a search for associated production of charginos and neutralinos
using a data set corresponding to an integrated luminosity of 2.3~\fbinv\ collected with
the D0 experiment during Run~II of the Tevatron proton-antiproton collider.
Final states containing three charged leptons and missing transverse energy are
probed for a signal from supersymmetry with four dedicated trilepton event selections.
No evidence for a signal is observed, and we set limits on the product of production
cross section and leptonic branching fraction. Within minimal
supergravity, these limits translate into bounds on \mzero\ and \mhalf\ that
are well beyond existing limits.

\end{abstract}

\maketitle

%\section{\label{sec:level1}First-level heading}
% sections are not used for PRL papers

Supersymmetry (SUSY) \cite{mssm} is one of the most popular extensions of 
the standard model (SM). SUSY can solve the hierarchy problem, 
allows the unification of gauge couplings, and provides a 
dark matter candidate. The analyses presented in this Letter are based on 
the supersymmetric extension 
of the SM with minimal field content, the so-called
minimal supersymmetric standard model (MSSM), which requires the addition of a 
SUSY partner for each SM particle, differing by half a unit in spin.
The supersymmetric partners of charged and neutral Higgs and gauge
bosons form two chargino ($\cha$) and four neutralino ($\chiz$)
mass eigenstates. 
Experiments at the CERN $e^+e^-$ Collider (LEP) have set lower limits on the
masses of SUSY particles. In particular, charginos
with mass lower than 103.5~GeV and sleptons ($\tilde{\ell}$) with mass  
below 95~GeV are excluded~\cite{lep_susy}. The results presented here are 
the extensions of an earlier search for charginos and neutralinos 
by the D0 collaboration based on 0.3~\fbinv\ of data~\cite{d0susy}.
The CDF collaboration has published limits for charginos and neutralinos 
using 2.0~\fbinv\ of data~\cite{cdfsusy}.

In $\ppbar$ collisions, charginos and neutralinos can be 
produced in pairs via an off-shell $W$ boson or the exchange of squarks.
They decay into fermions and the lightest neutralino~$\chiz_{1}$, which is assumed to
be the lightest supersymmetric particle (LSP) and to escape undetected.
This Letter describes the search for 
$\ppbar \rightarrow \cha_{1} \chiz_{2}$ in purely
leptonic decay modes in final states with missing transverse energy $\met$
and three
charged leptons ($e, \mu$ or $\tau$). This signature of three leptons can be particularly 
challenging in regions of parameter space where lepton momenta are very soft due to small
mass differences of the SUSY particles.
The analyses are based 
on $p\bar{p}$ collision data at a center-of-mass energy of 1.96~TeV recorded 
with the D0 detector at the Fermilab Tevatron Collider between March 2002 and June 2007 corresponding to an
integrated luminosity of 2.3~\fbinv, with the exception of the analysis 
using identified hadronic $\tau$ lepton decays, which is based on 1~\fbinv\ of data.

The D0 detector~\cite{d0det} consists of a central tracking system 
surrounded by a liquid-argon sampling calorimeter and a muon system. 
The inner tracking
systems, a silicon microstrip tracker and a central fiber tracker, reside in an axial magnetic field of 2~T. 
The $\eta$ coverage of the calorimeter extends
down to pseudorapidities of $|\eta| \approx 4$, where $\eta = -\ln\left[\tan(\theta/2)\right]$ and 
$\theta$ is the polar angle with respect to the proton beam direction. Muons are
identified in the inner tracking system as well as in the
outer muon system, which consists of three layers of 
tracking detectors and scintillator counters. An iron
toroidal magnet providing a field of 1.8~T is located between the two
innermost layers.  The muon system provides coverage for muon 
identification up to  $|\eta| \approx 2$.
A three-stage real-time trigger system reduces the total
rate from 2.5~MHz to about 100~Hz.
Events for the offline analyses are collected
by a combination of single lepton, di-lepton, and lepton plus track triggers.
Electrons and muons are selected by their specific energy deposition 
in the calorimeter and hits in the muon chambers, respectively. In addition,
high momentum tracks matched to the objects in the calorimeter and muon system
help to reduce the trigger rates.

Standard model and SUSY processes are simulated with the
event generators {\sc pythia}~\cite{pythia} (Drell-Yan, di-boson, $\Upsilon$, and 
$t\bar{t}$ events) 
and {\sc alpgen}~\cite{alpgen} ($W$+jet/$\gamma$ events). The simulation
of the detector geometry and response 
is based on {\sc geant}~\cite{geant}.
Detector noise and additional interactions are included using 
randomly triggered events recorded throughout the duration of the data-taking period.
The predictions for the SM backgrounds 
are normalized using the next-to-leading (NLO) and, for Drell-Yan
production, next-to-NLO theoretical cross sections, calculated
using  CTEQ6.1M parton distribution functions~\cite{cteq}.

The contributions from multijet background are estimated using
D0 data. For each analysis, samples dominated by 
multijet background are
defined that are identical to the search samples except for
reversed lepton identification requirements. In case of the electrons, jet-like electrons 
are selected based on the likelihood criterion 
(see below) while for the muons the isolation criteria (see below) are inverted. The normalization
of these samples is performed at an early stage of the selection
in a region of phase space that is dominated by 
multijet production.

The optimization of the analysis is done using 
minimal supergravity (mSUGRA)~\cite{nilles} as a reference model, in regions  
of parameter space with chargino, neutralino, and slepton masses ranging from 
100 to 200~GeV. The mSUGRA scenario can be described by five 
independent parameters: the unified scalar and gaugino masses \mzero\ and 
\mhalf, the ratio of the vacuum expectation values of the two Higgs 
doublets, $\tan\beta$, the unified trilinear coupling $A_0$, and the sign of
the Higgs mass parameter $\mu$. The SUSY spectra are calculated using {\sc softsusy}~\cite{softsusy}.  
The selection criteria are optimized to achieve the best average expected limit 
under the assumption that no signal is present in the data. A modified
frequentist approach~\cite{cls} is used to calculate limits at the 95\% C.L.
for each different final state and selection. 
Two choices of mSUGRA parameters (\mzero\ = 150 GeV and 
\mhalf\ = 250 (170) GeV, with $\tan\beta = 3$, $A_0 = 0$ and $\mu>0$) are 
used as a reference for a high-$p_T$ (low-$p_T$) signal,
labeled SUSY1 (SUSY2) in the plots shown in the following.

The reconstruction of isolated electrons exploits their characteristic energy deposition in the 
calorimeter.  All electromagnetic clusters with $|\eta| < 3.2$ are considered.
A track is required to point to the calorimeter energy 
cluster, and the track momentum and the calorimeter energy must be consistent. A likelihood 
discriminant is used to reject background contributions from jets, based on their differences 
in transverse and longitudinal shower shape as well as differences in isolation in the 
tracker. The selection of muons relies on a combination of tracks in the central tracker and 
pattern of hits in the muon detector within $|\eta| < 2.0$. Isolation criteria are imposed in both
the tracker and the calorimeter 
in order to suppress background contributions from jets. Two different type of muons, referred to as 
``loose'' and ``tight'', are used in the analyses. The classification of loose and tight muons 
depends on the level of calorimeter and tracker isolation of the candidate. The isolation in the 
calorimeter is based on the cell energies in a hollow cone of $0.1  < \Delta{\cal R} < 0.4$, where
$\Delta{\cal R} = \sqrt{(\Delta\eta)^2 + (\Delta\phi)^2}$. The tracker isolation is defined as 
the scalar sum of the transverse momenta of all tracks in a cone of $ \Delta{\cal R} < 0.4$ around
the muon track. The energies for both calorimeter and tracker isolation are required to be less 
than 4~GeV (2.5~GeV) for loose (tight) muons. Reconstruction efficiencies 
for both $e$ and $\mu$ are measured using $Z\to\ell\ell$ events, and  
the efficiencies in the Monte Carlo (MC) simulation are corrected for known differences 
according to the measurements in the data.

The reconstruction of hadronically decaying $\tau$ leptons is seeded by calorimeter clusters or 
tracks~\cite{d0-h-tautau} with $|\eta| < 2.5$. According to their signature in the detector, they are classified 
into three types. The signature of \ttype~1 (\ttype~2) consists of a single track with energy 
deposit in the hadronic (and the electromagnetic) calorimeter typically arising from 
$\pi^\pm$-like ($\rho^\pm$-like) decays. Three-prong decays (\ttype~3) are not considered here, since the 
background contribution from jets in this channel does not allow one to 
improve the sensitivity to a signal.
The separation of hadronic $\tau$ leptons and jets is based on a set of neural networks (NN),
one for each \ttype, 
exploiting the differences in longitudinal and transverse shower shapes as well as differences in 
the isolation in the calorimeter and the tracker~\cite{d0-h-tautau}. $Z\to\tau\tau$ MC events are 
used as the signal training sample for the neural networks, while multijet events from data serve as 
the background training sample. In order to ensure high efficiency for low $\tau$ lepton transverse 
momenta, the selection on the NN~output varies depending on the transverse momenta of the 
$\tau$ candidates to keep a constant efficiency of 60\%.
At a small rate, muons can be misidentified as one-prong hadronic $\tau$ lepton decays, 
and thus $\tau$ candidates to which a muon can be matched are rejected.

Jets are reconstructed with an iterative midpoint cone algorithm~\cite{jets} with cone radius 
of 0.5 and must be within $|\eta| < 2.5$. The \met\ is calculated from the vector sum of the transverse
components of the energy deposited in the
calorimeter cells and is corrected for electron, $\tau$ and jet energy calibrations as well
as the transverse momentum of muons.

In the following, four different channels are defined,
distinguished by the lepton content of the final state. For the di-electron
plus lepton channel (\eel) two identified electrons are required using
the electron identification criteria described above. In the di-muon plus lepton
channel (\mumul), one tight and one loose muon are required, while 
the selection in the electron, muon plus lepton channel (\emul) starts from one electron
and one tight muon. Finally, the muon, $\tau$ lepton plus lepton channel 
(\mutau) requires one tight muon and one hadronically decaying $\tau$ lepton
in the final state. In all cases, unless explicitly specified otherwise, the 
third lepton is reconstructed as an isolated track without using the standard lepton 
identification criteria.

For each of the \eel, \mumul\ and \emul\ channels, one ``low-$p_T$'' and 
one ``high-$p_T$'' selection is designed to exploit the different kinematic 
properties for various parameter points in the \mmplane. 
The \mutau\ channel is separated into two distinct selections based on 
the properties of the third 
object. One selection requires only an isolated track as third object,
as in the other three analyses (\mutaul\ selection). For the second selection, 
a fully reconstructed hadronic $\tau$ lepton is required ($\mu\tau\tau$
selection). Both $\mu\tau$ selections are identical over the whole \mmplane.

Each selection requires two identified leptons stemming
from the primary vertex with minimum 
transverse momenta of \ptone\ = 12~GeV and \pttwo\ = 8~GeV. Due to 
higher thresholds in the single muon triggers used for the
\mutau\ channel, the $p_T$ cut on the muon is tightened to 15~GeV
for this channel. If more than two leptons are identified that
satisfy the $p_{T}$ criteria, the two leptons with the highest $p_{T}$
are considered. In case of the \emul\ analysis, events are removed if two 
electrons or muons with an invariant mass compatible with that of the $Z$ boson
mass are found. This is called the preselection.
To further reduce the background, differences
in the kinematics and event topology compared to signal are exploited.
All selection criteria are summarized in Tables~\ref{tab:sel1} and \ref{tab:sel2}.

\begin{figure}
\epsfig{file=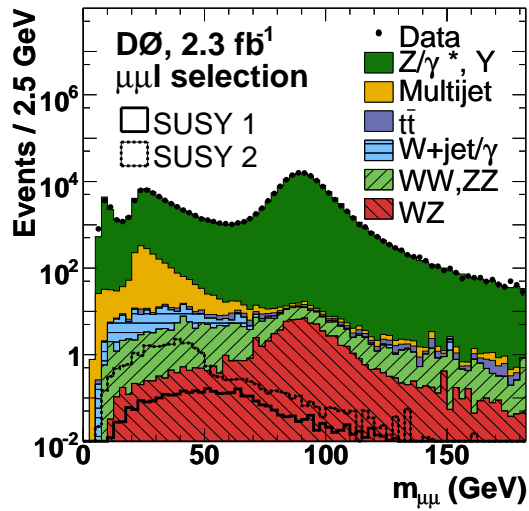,width=0.83\columnwidth,clip=true}
\caption{\label{fig:presel1}Invariant mass $m_{\mu\mu}$
(\mumul\ channel) for data (points), SM backgrounds
(shaded histograms), and SUSY signal (open histograms) after cut I (see
Table~\ref{tab:sel1}) for the low-$p_{T}$ selection.}
\end{figure}

The dominant background from Drell-Yan and $Z$ boson production in the \mumul\
and \eel\ channels as well as multijet background
can be reduced by selecting on the invariant mass $m_{\ell_{1}\ell_{2}}$
of the identified di-lepton system and the opening angle \dphi\ of the same two 
leptons in the transverse plane. As shown in Fig.~\ref{fig:presel1}, 
a major fraction of the di-lepton events from $Z$ boson decays can be rejected
by requiring the invariant mass $m_{\ell_{1}\ell_{2}}$ to be 
below the $Z$ resonance. A substantial fraction of the Drell-Yan events as well as 
the major part of events from multijet production are back-to-back
in the transverse plane and can be rejected by removing events
with large opening angle \dphi. 

\begin{figure}
\epsfig{file=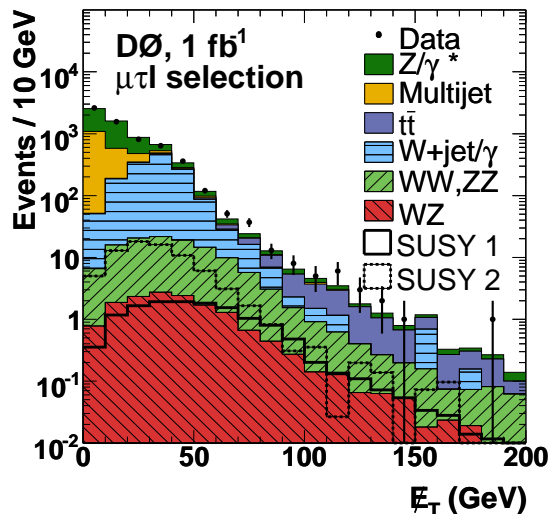,width=0.83\columnwidth,clip=true}
\caption{\label{fig:met1}Missing transverse energy \met\ 
($\mu\tau\ell$ selection) for data (points), SM backgrounds (shaded
histograms), and SUSY signal (open histograms) after cut I (see Table~\ref{tab:sel2}).}
\end{figure}

\begin{figure}
\epsfig{file=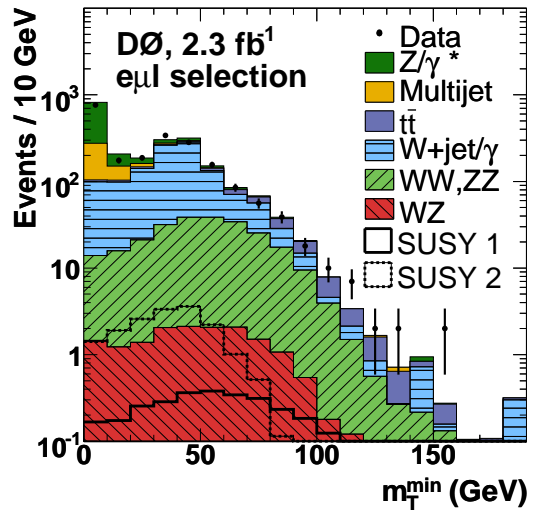,width=0.83\columnwidth,clip=true}
\caption{\label{fig:mtmin1}Minimum transverse mass $m_T^{\text{min}}$ 
(\emul\ channel) for data (points), SM backgrounds (shaded
histograms), and SUSY signal (open histograms) before applying the cut 
on $m_T^{\text{min}}$ (see Table~\ref{tab:sel1}) for the low-$p_{T}$ selection.}
\end{figure}

Another striking feature of the signal is the presence of large
\met\ due to the escaping neutralinos and neutrinos
in the final state. 
Thus selecting events with large \met\ is expected to further enhance
the signal, which is illustrated in Fig.~\ref{fig:met1}. 
However, backgrounds without true \met\ can potentially
satisfy this selection criterion, because of mismeasurements of the objects in the event or by failing to 
reconstruct them. If \met\ is caused by mismeasurement of an object, the
direction of the \met\ tends to be aligned with this object.  
For events with at least one jet, Sig(\met) is defined as
\begin{displaymath}
{\rm Sig}(\met) = \frac{\met}{\sqrt{\sum_{\text{jets}}\sigma^2(E_T^j||\met)}},
\end{displaymath}
where $\sigma(E_T^j||\met)$ is the jet energy resolution projected on the \met\ direction.
As a result, Sig(\met) is expected to be small for events with poorly measured jets. 
Rejecting events with small minimal transverse mass 
$m_T^{\text{min}} = \text{min}(m_T^{\ell_1}, m_T^{\ell_2})$, where
$m_T^{\ell} = \sqrt{2 p_T^{\ell}\met\left[1-\cos\Delta\phi(\ell,\met)\right]}$,
removes events with mismeasured leptons as illustrated in 
Fig.~\ref{fig:mtmin1}. 
Other events with large jet activity, in particular $t\bar{t}$ production, can be removed
with a cut on $H_T$, the scalar sum of the $p_{T}$ of all jets with $p_{T} > 15$~GeV.

\begin{figure}
\epsfig{file=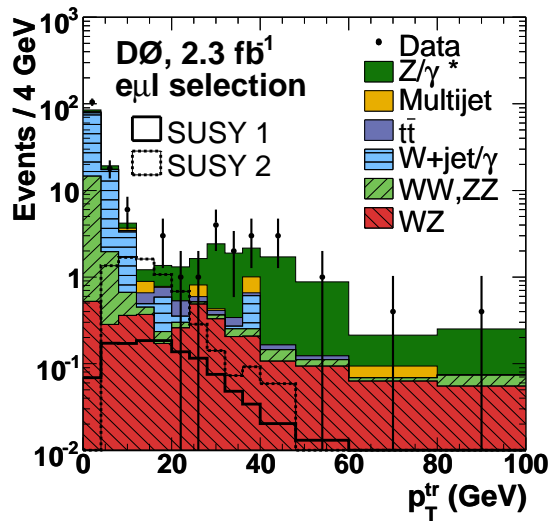,width=0.83\columnwidth,clip=true}
\caption{\label{fig:track}Transverse momentum of the track 
(\emul\ channel) for data (points), SM backgrounds (shaded
histograms), and SUSY signal (open histograms) after all \met\ related
cuts are applied (cut III, see Table~\ref{tab:sel1}) for the
low-$p_{T}$ selection.}
\end{figure}

\begin{figure}
\epsfig{file=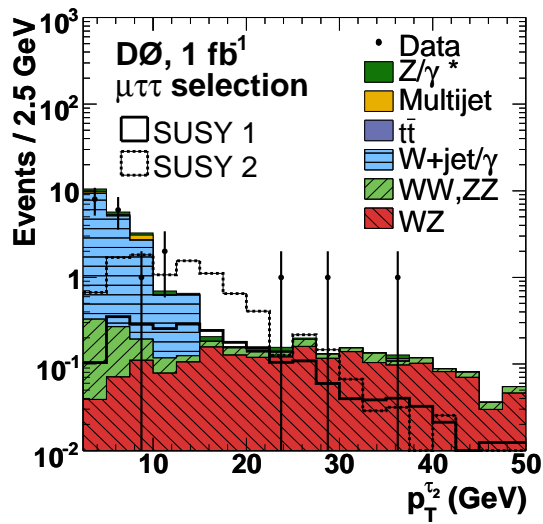,width=0.83\columnwidth,clip=true}
\caption{\label{fig:taupt1}Transverse momentum of the second $\tau$ lepton candidate
($\mu\tau\tau$ selection) for data (points), SM
backgrounds (shaded histograms), and SUSY signal (open histograms)
after cut III (see Table~\ref{tab:sel2}).}
\end{figure}

Unlike most SM backgrounds, signal events contain three charged leptons.
This can be exploited to remove most of the remaining background, which
is dominated by $W$+jet production at this stage of the selection.
The \eel, \mumul, \emul, and \mutaul\ selections only require an additional track
that must be isolated in both the tracking system and the calorimeter as 
indication of this third lepton. Dropping the lepton identification criteria in this case
increases the signal efficiency and includes all three lepton flavors
in the selection. The distribution of the transverse momentum of this
additional track is presented in Fig.~\ref{fig:track} 
after \met, Sig(\met ) and $m_T^{\text{min}}$ cuts are applied. Selection of tracks with high transverse 
momentum clearly enhances signal over background.
For the $\mu\tau\tau$ channel, a well-identified second $\tau$ lepton is required 
instead of the track. Since the $\tau$ 
lepton selection imposes different criteria than the track 
selection, some signal loss due to the third track criterion can be regained
using this selection. In particular at high $\tan\beta$, this selection is 
favored, since most of the leptons in the final state are expected to be
$\tau$ leptons. Figure~\ref{fig:taupt1} shows the distribution of the transverse
momentum for the second $\tau$ lepton candidate.

After the third object selection, the remaining background consists
mainly of $W$ 
and $Z$ boson as well as di-boson production. These backgrounds 
are addressed in the following. The remaining $Z$ boson background mainly 
consists of events where one of the leptons from the $Z$ boson decay is 
not reconstructed in the calorimeter or muon system, but instead 
a jet or photon from initial or final state radiation is misidentified
as one of the two initially selected leptons. However, the missed lepton
from the $Z$ boson decay is then selected as the third track. This unique feature 
provides two handles to reject this background. Due to the non-reconstruction
of one of the leptons, the \met\ tends to point into the direction of the 
track. Thus the transverse mass calculated from the track and \met\
should be small due to the small opening angle 
$\Delta\phi_{tr,\not \hspace{-0.4mm} \scriptsize{E_T}}$. In addition, 
the invariant mass of the track and one of the leptons, \mll, is expected
to be consistent with the $Z$ boson mass. The same is true for $WZ$ production,
where again one of the leptons from the $Z$ decay is reconstructed in
the tracking system. 

For $W$ boson production, only one real lepton is expected from the
decay of the $W$ boson, the second lepton is mimicked by a jet or
a photon. In the case of jets, the identification criteria for that lepton
tend to be of worse quality, while in case of photon conversions, no hits
in the innermost layers of the tracking detector are expected for 
the track corresponding to the converted photon. Thus, requiring 
high quality leptons (tight likelihood for electrons and 
very tight track isolation for muons) or hits in the first two layers of the tracking system
is expected to reduce $W$+jet/$\gamma$ background. 
To keep signal efficiencies high, these requirements are only used
if the event properties and kinematics are similar to expectations from 
$W$ boson production (see Table~\ref{tab:sel1}). 
In case of the \mutaul\ selection, a dedicated likelihood discriminant
is developed to remove the background from $W$ boson production. This
likelihood uses the transverse masses calculated for all of the three
leptons as well as products of \met\ and lepton transverse momenta.
In case of the  $\mu\tau\tau$ selection, the product of the two NN outputs 
for $\tau$ lepton identification is used to remove events containing 
misidentified $\tau$ candidates.

Finally, the  different event kinematics for signal and background are exploited
to obtain better signal sensitivity.
Since background is expected to have low transverse momentum for 
the third track or small \met, a cut on the product of track $p_T$ and \met\
effectively rejects any remaining background contributions. 
In addition, the vectorial sum of the lepton transverse momenta and \met\ 
should equal the transverse momentum of the third track in case of signal
events. Thus the $p_T$ balance
\begin{displaymath}
p_{T}^{\text{bal}} = \frac{|\vec{p}_T^{\ \ell_1} + \vec{p}_T^{\ \ell_2} + \not \hspace{-0.5mm} \vec{E}_T|}{p_T^{tr}}
\end{displaymath}
is expected to peak at 1 for a signal, while for background a broad distribution is 
expected.

\begin{figure}
\epsfig{file=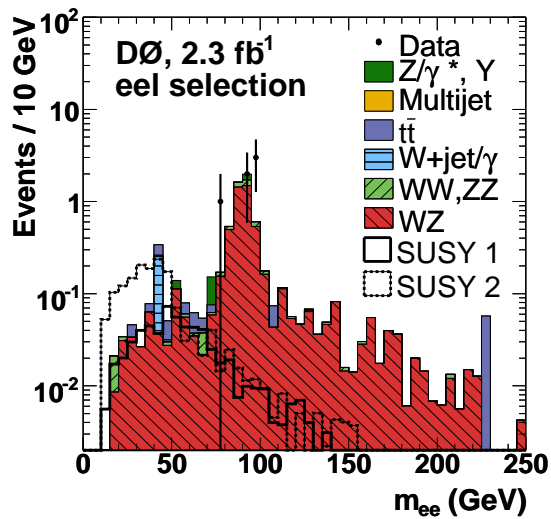,width=0.83\columnwidth,clip=true}
\caption{\label{fig:wz}Distribution of the invariant mass $m_{ee}$ 
(\eel\ channel) for data (points), SM backgrounds (shaded histograms),
and SUSY signal (open histograms) with all cuts applied except the
$m_{ee}$ requirement for the low-$p_{T}$ selection.}
\end{figure}

After all selection criteria are applied, the expected background is dominated by 
irreducible background from 
$WZ$ production, as is evident from the marginal distribution of the di-electron 
invariant mass in the \eel\ selection shown in Fig.~\ref{fig:wz}.
A detailed comparison of background expectation and events observed in
data together with efficiency expectations from a typical SUSY signal are shown in 
Tables~\ref{tab:comp1} and \ref{tab:comp2} for the low-$p_T$ and
high-$p_T$ selection, respectively, while Table~\ref{tab:comp3}
presents the results for the $\mu\tau$ selections. In general, good agreement between
data and expectation from SM processes is observed. Combining
all low-$p_T$ and $\mu\tau$ selections, a background of 
$5.4\pm 0.4({\rm stat})\pm 0.4({\rm syst})$ events from SM processes 
is expected with 9 events observed in the data. The probability to observe 9 or more events in the data given the expected background
is 10\%. The expectation for
the reference signal point SUSY2 is $9.3\pm 0.3({\rm stat})\pm
0.8({\rm syst})$ events. The high-$p_T$ selection yields
$3.3\pm 0.3({\rm stat})\pm 0.3({\rm syst})$ events from SM processes, while 
4 events are observed in data. The expected reference signal for
parameter point SUSY1 is $0.9\pm 0.1({\rm stat})\pm
0.1({\rm syst})$ events.

%%%%%%%%%%%%%%%%%%%%%%%%%%%%%%%%%%%%%%%%%%%%%%%%%%%%%%%%%%%%%%%%%%%%%%%%%
\begin{table*}
\caption{\label{tab:sel1}Selection criteria for the \mumul, \eel\ and \emul\ 
analyses (all energies, masses and momenta in GeV, angles in radians) for the 
low-$p_T$ selection and high-$p_T$ selection,  see text for further details.} 
\begin{ruledtabular}
\renewcommand{\arraystretch}{1.2}
\begin{tabular}{cccccccc}
 & Selection  & \multicolumn{2}{c}{\mumul} 
              & \multicolumn{2}{c}{\eel} 
              & \multicolumn{2}{c}{\emul}\\
 &            & low $p_T$ & high $p_T$ & low $p_T$ & high $p_T$ & low $p_T$ & high $p_T$\\
\hline
I & \ptone, \pttwo & 
  $>$12, $>$8 & $>$18, $>$16 & $>$12, $>$8 & $>$20, $>$10 & $>$12, $>$8\footnotemark[1] & $>$15, $>$15\\
\hline
%Dilepton Cuts
 & $m_{\ell_{1}\ell_{2}}$\footnotemark[2] &$\in [20,60]$ & $\in [0,75]$ &$\in [18,60]$ &$\in [0,75]$ & -- & --\\
II & $\Delta\phi_{\ell_{1}\ell_{2}}$ &$<$2.9 & $<$2.9 &$<$2.9 &$<$2.9 & -- & --\\
\hline
% Missing $E_T$
 & \met\ & 
 $>$20 & $>$20 &$>$22 & $>$20 & $>$20 & $>$20 \\
 & Sig(\met) & 
 $>$8 & $>$8 & $>$8 &$>$8 & $>$8 & $>$8 \\
 & $m_T^{\mathrm{min}}$ &
 $>$20 & $>$20 &$>$20 & $>$14 & $>$20 & $>$15 \\
III & jet-veto $H_T$ &-- & $<$80 & -- & -- & -- & --\\
\hline 
% Third Lepton
IV & \pttr &
 $>$5 & $>$4 &$>$4  &$>$12 & $>$6 & $>$6\\
\hline
% m_ltr, METxpt, sum/pt
  & $m_T^{tr}$ & $>$10 & $>$10 & $>$10& $>$10 &$>$10 & $>$8\\
V & \mll & $\notin [80,110]$ & -- &-- &-- & $<$70 &  $<$70 \\ 
\hline
VI &  anti $W$ & -- & -- &
tight likelihood\footnotemark[3] & -- & \multicolumn{2}{c}{tight likelihood\footnotemark[4]}\\
 & & & & & & \multicolumn{2}{c}{hit in 2 inner layers\footnotemark[4]}\\
 & & & & & & \multicolumn{2}{c}{very tight muon isolation\footnotemark[5]}\\
 & & & & & & \multicolumn{2}{c}{$\sum_{0.05<\Delta{\cal
       R}<0.4} p_{T}^{\text{track}}<$1}\\
\hline
 & \met\ $\times$ \pttr & 
 $>$200 &$>$300 & $>$220 & -- & -- & --\\
VII & $p_{T}^{\text{bal}}$ & $<$4 & $<$4& $<$4& $<$4&$<$2 &$<$2 \\
\end{tabular}
\renewcommand{\arraystretch}{1.0}
\end{ruledtabular}
\footnotetext[1]{\ptone\ and \pttwo\ are electron and muon $p_T$, respectively.}
\footnotetext[2]{$\ell$ refers to the two identified leptons}
\footnotetext[3]{for \pttr\ $<$15 GeV}
\footnotetext[4]{for $m_T^\mu$ $\in [40,90]$ GeV}
\footnotetext[5]{for $m_T^e$ $\in [40,90]$ GeV}
\end{table*}
%%%%%%%%%%%%%%%%%%%%%%%%%%%%%%%%%%%%%%%%%%%%%%%%%%%%%%%%%%%%%%%%%%%%%%%%%%

%%%%%%%%%%%%%%%%%%%%%%%%%%%%%%%%%%%%%%%%%%%%%%%%%%%%%%%%%%%%%%%%%%%%%%%%%%
\begin{table}
\caption{\label{tab:sel2}Criteria for the \mutaul\ and $\mu\tau\tau$ 
selections (all energies, masses and momenta in GeV, angles in radians),
see text for further details.} 
\begin{ruledtabular}
\renewcommand{\arraystretch}{1.2}
\begin{tabular}{ccccc}
 & Selection  & \mutaul 
              & 
              & $\mu\tau\tau$\\
\hline
I & \ptone, \pttwo & 
  \multicolumn{3}{c}{$>$15, $>$8\footnotemark[1]}\\
\hline
%Dilepton Cuts
II & $\Delta\phi_{\ell_{1}\ell_{2}}$ & \multicolumn{3}{c}{$<$2.9}\\
\hline
% Missing $E_T$
 & \met\ &  \multicolumn{3}{c}{$>$20}\\
 & Sig(\met) & \multicolumn{3}{c}{$>$8} \\
 & $m_T^{\mu}$ & \multicolumn{3}{c}{$>$20}\\
III & jet-veto $H_T$ & \multicolumn{3}{c}{$<$80}\\
\hline 
% Third Lepton
IV & \pttr &
 $>$3 & $p_T^{\tau_2}$ & $>$4\\
\hline
% m_ltr, METxpt, sum/pt
  & $\Delta\phi_{tr,\not \hspace{-0.4mm} \scriptsize{E_T}}$ & $>$0.5 & $\Delta\phi_{\tau_2,\not \hspace{-0.4mm} \scriptsize{E_T}}$ & $>$0.5\\
V & \mll & $<$60 & & $<$60\\ 
\hline
 &  anti $W$ & likelihood & & likelihood\\
VI & & & $NN_{\tau_1}\times NN_{\tau_2}$ & $>$0.7\\
\hline
VII &\met\ $\times$ \pttr & $>$300 & $p_{T}^{\text{bal}}$ & $<$3.5\\
\end{tabular}
\renewcommand{\arraystretch}{1.0}
\end{ruledtabular}
\footnotetext[1]{\ptone\ and \pttwo\ are muon and $\tau$ lepton $p_T$, respectively.}
\end{table}
%%%%%%%%%%%%%%%%%%%%%%%%%%%%%%%%%%%%%%%%%%%%%%%%%%%%%%%%%%%%%%%%%%%%%%%%%%

%%%%%%%%%%%%%%%%%%%%%%%%%%%%%%%%%%%%%%%%%%%%%%%%%%%%%%%%%%%%%%%%%%%%%%%%%
\begin{table*}
\caption{\label{tab:comp1}Numbers of events observed in data and
expected for background and reference signal efficiency (SUSY2, see text) in percent at various
stages of the selection with statistical uncertainties for the low-$p_T$ selection. Each
row corresponds to a group of cuts, as detailed in
Table~\ref{tab:sel1}. 
}
\begin{ruledtabular}
\begin{tabular}{ccrr@{$\pm$}lr@{$\pm$}lcrr@{$\pm$}lr@{$\pm$}lcrr@{$\pm$}lr@{$\pm$}l}
Selection & & \multicolumn{5}{c}{\mumul} &
              & \multicolumn{5}{c}{\eel} &
              & \multicolumn{5}{c}{\emul}\\
 & & Data & \multicolumn{2}{c}{Backgrd.} & \multicolumn{2}{c}{Eff. (\%)} &
 & Data & \multicolumn{2}{c}{Backgrd.} & \multicolumn{2}{c}{Eff. (\%)} &
 & Data & \multicolumn{2}{c}{Backgrd.} & \multicolumn{2}{c}{Eff. (\%)}\\
\hline
% lepton ID, pt1, pt2
I & \hspace*{10mm} & 194006 & 195557 & 177 & 19.9 & 0.3 & \hspace*{10mm}
  & 235474 & 232736 & 202 & 15.5 & 0.2 & \hspace*{10mm}
  & 16630 & 16884 & 75 & 10.5 & 0.1\\
II & & 22766 & 26067 & 88 & 14.6 & 0.2 & 
  &  31365& 27184 & 64 & 11.0 & 0.2 &
  &  \multicolumn{5}{c}{}\\
III & & 178  & 181 & 6.4 & 8.8 & 0.1 &
  &  515& 512 & 12 & 6.8 & 0.2 &
  &  1191& 1177 & 20 & 5.8 & 0.1\\
IV & & 7 & 2.9 & 0.7 & 3.4 & 0.1 &
  &  16& 9.3 & 2.0 & 3.0 & 0.1 &
  &  22& 18.0 & 1.2 & 2.4 & 0.1\\
V & & 4 & 2.2 & 0.5 & 3.0 & 0.1 & 
  & 9 & 5.9 & 1.7 & 2.5 & 0.1 &
  & 3 & 3.5 & 0.5 & 2.0 & 0.1\\
VI & & \multicolumn{5}{c}{} &
  & 6 & 3.1 & 0.4 & 2.2 & 0.1 &
  & 2 & 1.6 & 0.4 & 1.5 & 0.1\\
VII & & 4 & 1.2 & 0.2 & 2.8 & 0.1 &
  & 2 & 1.8 & 0.2 & 2.1 & 0.1 &
  & 2 & 0.8 & 0.2 & 1.3 & 0.1\\
\end{tabular}
\end{ruledtabular}
\end{table*}

%%%%%%%%%%%%%%%%%%%%%%%%%%%%%%%%%%%%%%%%%%%%%%%%%%%%%%%%%%%%%%%%%%%%%%%%%
\begin{table*}
\caption{\label{tab:comp2}Numbers of events observed in data and
expected for background and reference signal efficiency (SUSY1, see text) in percent at various
stages of the selection with statistical uncertainties for the high-$p_T$ selection. Each
row corresponds to a group of cuts, as detailed in
Table~\ref{tab:sel1}.
}
\begin{ruledtabular}
\begin{tabular}{ccrr@{$\pm$}lr@{$\pm$}lcrr@{$\pm$}lr@{$\pm$}lcrr@{$\pm$}lr@{$\pm$}l}
Selection & & \multicolumn{5}{c}{\mumul} &
              & \multicolumn{5}{c}{\eel} &
              & \multicolumn{5}{c}{\emul}\\
 & & Data & \multicolumn{2}{c}{Backgrd.} & \multicolumn{2}{c}{Eff. (\%)} &
 & Data & \multicolumn{2}{c}{Backgrd.} & \multicolumn{2}{c}{Eff. (\%)} &
 & Data & \multicolumn{2}{c}{Backgrd.} & \multicolumn{2}{c}{Eff. (\%)}\\
\hline
% lepton ID, pt1, pt2
I & \hspace*{10mm} & 140417 & 141781 & 120 & 19.6 & 0.2 & \hspace*{10mm}
  & 171001 & 170197 & 175 & 18.1 & 0.2 & \hspace*{10mm}
  & 4617 & 4709 & 23 & 11.5 & 0.2\\
II & & 10349 & 10645 & 51 & 15.3 & 0.2 &
  &  8273& 7937 & 39 & 12.8 & 0.1 &
  & \multicolumn{5}{c}{}\\
III & & 173 & 176 & 5.7 & 11.4 & 0.2 &
  & 244 & 264 & 10 & 10.8 & 0.1&
  & 727 & 738 & 11 & 8.9 & 0.1\\
IV & & 7  & 3.8 & 0.5 & 5.9 & 0.1&
  & 0 & 1.5 & 0.3 & 4.0 & 0.1 &
  & 11 & 12.7 & 0.9 & 4.1 & 0.1\\
V & & 4 & 2.9 & 0.4 & 5.5 & 0.1 &
  & 0 & 1.1 & 0.3 & 3.6 & 0.1 &
  & 2 & 2.8 & 0.5 & 2.9 & 0.1\\
VI & & \multicolumn{10}{c}{} &
 &  & 0 & 1.0 & 0.2 & 2.4 & 0.1\\
VII & & 4 & 2.0 & 0.3 & 5.0 & 0.1 &
  & 0 & 0.8 & 0.1 & 3.6 & 0.1 &
  & 0 & 0.5 & 0.1 & 2.1 & 0.1\\
\end{tabular}
\end{ruledtabular}
\end{table*}

%%%%%%%%%%%%%%%%%%%%%%%%%%%%%%%%%%%%%%%%%%%%%%%%%%%%%%%%%%%%%%%%%%%%%%%%%
\begin{table*}
\caption{\label{tab:comp3}Numbers of events observed in data and
expected for background and reference signal efficiency (SUSY2 for the $\mu\tau\ell$ selection and
SUSY1 for the $\mu\tau\tau$ selection, see text) in percent at various
stages of the selection with statistical uncertainties for the
$\mu\tau$ selections. Each
row corresponds to a group of cuts, as detailed in
Table~\ref{tab:sel2}. 
}
\begin{ruledtabular}
\begin{tabular}{cccrr@{$\pm$}lr@{$\pm$}lcrr@{$\pm$}lr@{$\pm$}lc}
& Selection & & \multicolumn{5}{c}{$\mu\tau\ell$} & 
 & \multicolumn{5}{c}{$\mu\tau\tau$} &\\
 & & & Data & \multicolumn{2}{c}{Backgrd.} & \multicolumn{2}{c}{Eff. (\%)} &
 & Data & \multicolumn{2}{c}{Backgrd.} & \multicolumn{2}{c}{Eff. (\%)}  &\\
\hline
% lepton ID, pt1, pt2
\hspace*{10mm} & I & \hspace*{20mm}& 6251 & 6238&30 & 8.1&0.2 & \hspace*{40mm}
  & 6251 & 6238 & 30 & 12.4 & 0.2 &\hspace*{10mm} \\
& II & & 3473 & 3416 & 17 & 6.9 & 0.2 &
  & 3473 & 3416 & 17 & 10.8 & 0.2 & \\
& III & & 1180& 1154 & 14 & 4.5 & 0.1 &
  & 1180 & 1154 & 14 & 8.7 & 0.1 & \\
& IV & & 103& 110.0 & 5.1 & 2.9 & 0.1 &
  & 20 & 22.6 & 2.6 & 2.2 & 0.1 & \\
& V & & 67 & 52.8 & 4.1 & 2.1 & 0.1 &
  & 7 & 8.0 & 1.5 & 1.7 & 0.1 & \\
& VI & & 4 & 2.9 & 0.4 & 1.5 & 0.1 &
  & 3 & 1.9 & 0.5 & 1.4 & 0.1 & \\
& VII & & 0 & 0.8 & 0.1 & 1.2 & 0.1 &
  & 1 & 0.8 & 0.2 & 1.3 & 0.1 & \\
\end{tabular}
\end{ruledtabular}
\end{table*}

The estimate of the expected numbers of signal and background events depends
on various measurements with associated systematic uncertainties:
integrated luminosity (6\%) \cite{lumi}, trigger efficiency, lepton identification and 
reconstruction efficiencies (4\%), jet and $\tau$ energy calibration in signal 
(2\%--6\%) and background events (2\%--9\%), PDF uncertainties (3\%--4\%), 
and modeling of the multijet background (2\%--30\%). All uncertainties, except 
the last one, are correlated among the different channels.

No evidence for a signal is observed. The search results can be translated
into upper limits on the product of cross section and branching fraction into
three charged leptons,~\sigbr. Limits are based on the combination of all low-
and high-$p_T$ selections. Events appearing in multiple analyses are uniquely
assigned to the channel with the best signal to background ratio. Correlated
systematic uncertainties are taken into account.

To calculate the limits, the mass relations between the particles involved in the 
decay chain of chargino and neutralino have to be known. The 
mSUGRA model is used to calculate the mass differences between \chaone, \chiztwo, and 
\chizone, which approximately corresponds to the assumption 
\mcha$\approx$\mchitwo$\approx$2\mchione. For slepton
and sneutrino masses, 
several scenarios are taken into account.

\begin{figure}
\epsfig{file=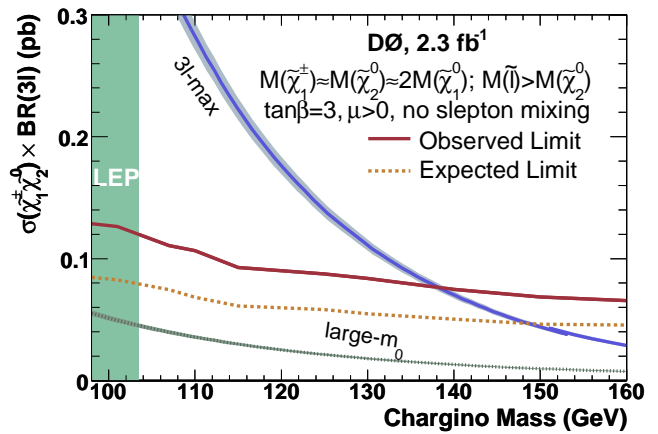,width=0.98\columnwidth,clip=true}
\caption{\label{fig:limit_mcha} Upper limit at the 95\% C.L. on \sigbr\ as a function of
\chaone\ mass, in comparison with 
the expectation for two SUSY scenarios (see text). PDF and
renormalization/factorization scale
uncertainties on the predicted cross section are shown as shaded bands.}
\end{figure}

Figure~\ref{fig:limit_mcha} shows the limit on \sigbr\ as a function of chargino mass 
assuming that sleptons and 
sneutrinos are heavier than the lightest chargino and the second-lightest 
neutralino, and assuming that slepton mixing can be neglected. In this case, 
both \chaone\ and \chiztwo\ decay via three-body decays, and branching fractions do not 
depend on the lepton flavor. The limit is compared with the NLO cross 
section~\cite{prospino} multiplied by branching fractions calculated in 
the limit of heavy sleptons (``large-$m_0$'' scenario) and for slepton 
masses just above the mass of the \chiztwo, in which case the leptonic branching 
fraction for three-body decays is maximized (``3l-max'' scenario).
For the latter, an observed (expected) lower limit at the 95\%~C.L. on the chargino mass 
is set at 138~GeV (148~GeV).

\begin{figure}
\epsfig{file=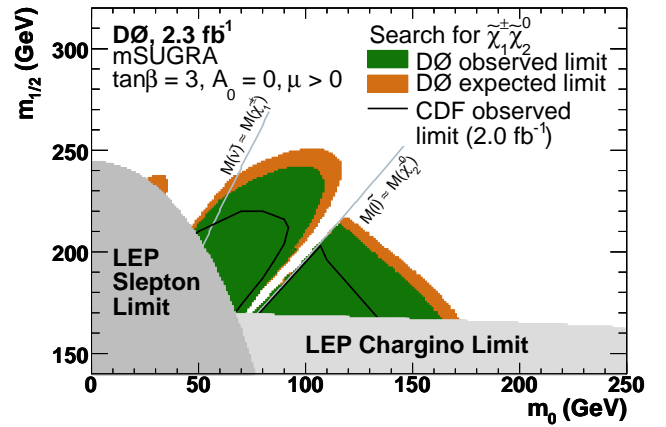,width=0.98\columnwidth,clip=true}
\caption{\label{fig:limit_m0m12} Region in the \mmplane\ excluded by
  the combination of the D0 analyses (green), by LEP searches for
  charginos (light grey) and sleptons (dark grey)
  \cite{lep_susy} and CDF (black line) \cite{cdfsusy}. 
 The assumed mSUGRA parameters are $\tan\beta = 3$, $A_0 = 0$ and $\mu > 0$. 
}
\end{figure}

Alternatively, the results can be interpreted within mSUGRA. 
To obtain the efficiency for any given point in the \mmplane,
selection efficiencies are first determined separately for three-body
decays of chargino and neutralino as well as two-body decays via sleptons
and sneutrinos. The variation of these efficiencies throughout the
plane can then be parametrized for each selection as a function of the
chargino, slepton and sneutrino masses. Using the mSUGRA prediction of
branching fractions and masses~\cite{pythia} \cite{softsusy}
\cite{sdecay}, 
the parametrized efficiencies
are used to calculate the total efficiency for each point in the
\mmplane.
Figure~\ref{fig:limit_m0m12} shows the region excluded in the \mmplane\ 
for $\tan\beta = 3$, $A_0 = 0$ and $\mu > 0$ in comparison with the limits from chargino and 
slepton searches at LEP~\cite{lep_susy} and CDF~\cite{cdfsusy}. 
The shape of the excluded region is driven by the relation of gaugino and slepton masses 
throughout the plane, which affects the branching fraction into three charged leptons as 
well as the efficiencies of the selections. For slepton masses just below the \chiztwo\
mass, one of the leptons from the \chiztwo\ decay has very small momentum, rendering 
the trilepton selections inefficient. For sneutrinos lighter than the \chaone\ and 
\chiztwo, two-body decays into sneutrinos open up, leading to a smaller branching fraction 
into three charged leptons as well as a reduced selection efficiency due to the 
small mass difference between sneutrino and chargino. For the
intermediate region at $m_{1/2} \approx 245$~GeV, 
chargino decays via $W$~bosons compete with decays via sleptons, leading to a 
reduction in leptonic branching fraction with increasing \mhalf\ both below and above 
the threshold for production of a real $W$ boson.

\begin{figure}
\epsfig{file=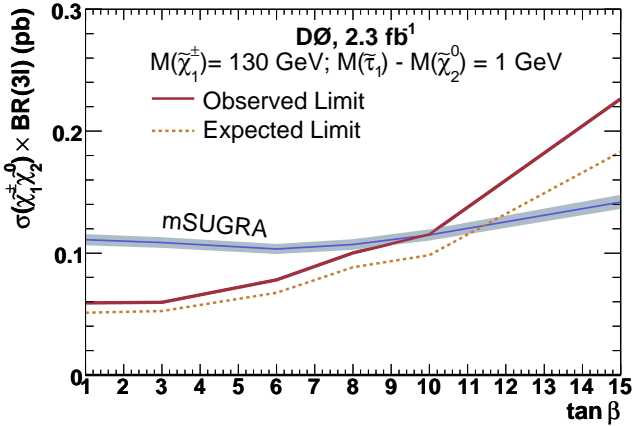,width=0.98\columnwidth,clip=true}
\caption{\label{fig:limit_tanb} Upper limit at the 95\% C.L. on \sigbr\ as a function of \tanb\ in 
comparison with the prediction for a chargino mass of 130~GeV and
$m_{\tilde{\tau}} - m_{\tilde{\chi}^{0}_{2}}$ = 1~GeV.}
\end{figure}

The excluded region in the \mmplane\ depends on the choice of \tanb, as the branching 
fraction into $\tau$ leptons increases as a function of \tanb. Figure~\ref{fig:limit_tanb}
shows the limit on \sigbr\ as a function of \tanb\ for a chargino mass of 130~GeV 
and fixing \mzero\ such that the lightest stau ($\tilde{\tau_{1}}$) is heavier than the \chiztwo\ by 
1~GeV. The latter choice results in three-body decays with maximal leptonic branching 
fraction. The leptonic branching fraction into three $\tau$ leptons increases as
a function of \tanb, 
reaching values above 50\% for $\tan\beta > 15$. Because all selections 
have been designed to 
be efficient for $\tau$ leptons, the limit remains 
stable within a factor of two for $\tan\beta \lesssim 10$, allowing one to 
exclude charginos with a mass of 130~GeV up to \tanb\ of 9.6.

To summarize, a data set collected with the D0 detector corresponding to an 
integrated luminosity of 2.3~\fbinv\ has been analyzed in search of the associated 
production of charginos and neutralinos in final states with three charged 
leptons and \met. No evidence for a signal is observed, 
and upper limits on the product of production cross section and leptonic branching 
fraction have been set. Within the reference model of mSUGRA with 
$\tan\beta = 3$, $A_0 = 0$, and $\mu > 0$, this result 
translates into excluded regions in the \mmplane\ that significantly
extend beyond all existing limits from direct searches for supersymmetric particles.

% acknowledgement_paragraph_r2.tex                         11/25/08
%
We thank the staffs at Fermilab and collaborating institutions, 
and acknowledge support from the 
DOE and NSF (USA);
CEA and CNRS/IN2P3 (France);
FASI, Rosatom and RFBR (Russia);
CNPq, FAPERJ, FAPESP and FUNDUNESP (Brazil);
DAE and DST (India);
Colciencias (Colombia);
CONACyT (Mexico);
KRF and KOSEF (Korea);
CONICET and UBACyT (Argentina);
FOM (The Netherlands);
STFC (United Kingdom);
MSMT and GACR (Czech Republic);
CRC Program, CFI, NSERC and WestGrid Project (Canada);
BMBF and DFG (Germany);
SFI (Ireland);
The Swedish Research Council (Sweden);
CAS and CNSF (China);
and the
Alexander von Humboldt Foundation (Germany).
%
   % input acknowledgement

\end{document}